
\input epsf
%
%
\catcode`\@=11 
\newcount\yearltd\yearltd=\year\advance\yearltd by -1900
%

\def\draftmode{\message{ DRAFTMODE }\def\draftdate{{\rm preliminary draft:
\number\month/\number\day/\number\yearltd\ \ \hourmin}}%
\headline={\hfil\draftdate}\writelabels\baselineskip=20pt plus 2pt minus 2pt
 {\count255=\time\divide\count255 by 60 \xdef\hourmin{\number\count255}
  \multiply\count255 by-60\advance\count255 by\time
  \xdef\hourmin{\hourmin:\ifnum\count255<10 0\fi\the\count255}}}
\def\nolabels{\def\wrlabeL##1{}\def\eqlabeL##1{}\def\reflabeL##1{}}
\def\writelabels{\def\wrlabeL##1{\leavevmode\vadjust{\rlap{\smash%
{\line{{\escapechar=` \hfill\rlap{\sevenrm\hskip.03in\string##1}}}}}}}%
\def\eqlabeL##1{{\escapechar-1\rlap{\sevenrm\hskip.05in\string##1}}}%
\def\reflabeL##1{\noexpand\llap{\noexpand\sevenrm\string\string\string##1}}}
\nolabels
%
\global\newcount\secno \global\secno=0
\global\newcount\meqno \global\meqno=1
\def\newsec#1{\global\advance\secno by1\message{(\the\secno. #1)}
\global\subsecno=0\eqnres@t\noindent{\bf\the\secno ~#1}
\writetoca{{\secsym} {#1}}\par\nobreak\medskip\nobreak}
\def\eqnres@t{\xdef\secsym{\the\secno.}\global\meqno=1\bigbreak\bigskip}
\def\sequentialequations{\def\eqnres@t{\bigbreak}}\xdef\secsym{}
\global\newcount\subsecno \global\subsecno=0
\def\subsec#1{\global\advance\subsecno by1\message{(\secsym\the\subsecno. #1)}
\ifnum\lastpenalty>9000\else\bigbreak\fi
\noindent{\it\secsym\the\subsecno ~#1}\writetoca{\string\quad
{\secsym\the\subsecno.} {#1}}\par\nobreak\medskip\nobreak}
\def\appendix#1{\global\meqno=1\global\subsecno=0\xdef\secsym{\hbox{#1}}
\bigbreak\bigskip\noindent{\bf #1}
\writetoca{{#1}}\par\nobreak\smallskip\nobreak}
%
%
\def\eqnn#1{\xdef #1{(\secsym\the\meqno)}\writedef{#1\leftbracket#1}%
\global\advance\meqno by1\wrlabeL#1}
\def\eqna#1{\xdef #1##1{\hbox{$(\secsym\the\meqno##1)$}}
\writedef{#1\numbersign1\leftbracket#1{\numbersign1}}%
\global\advance\meqno by1\wrlabeL{#1$\{\}$}}
\def\eqn#1#2{\xdef #1{(\secsym\the\meqno)}\writedef{#1\leftbracket#1}%
\global\advance\meqno by1$$#2\eqno#1\eqlabeL#1$$}
%
%
\global\newcount\refno \global\refno=1
\newwrite\rfile
\def\ref{$^{\the\refno}$\nref}
\def\nref#1{\xdef#1{\the\refno.}\writedef{#1\leftbracket#1}%
\ifnum\refno=1\immediate\openout\rfile=refs.tmp\fi
\global\advance\refno by1\chardef\wfile=\rfile\immediate
\write\rfile{\noexpand\item{#1\ }\reflabeL{#1\hskip.31in}\pctsign}\findarg}
\def\findarg#1#{\begingroup\obeylines\newlinechar=`\^^M\pass@rg}
{\obeylines\gdef\pass@rg#1{\writ@line\relax #1^^M\hbox{}^^M}%
\gdef\writ@line#1^^M{\expandafter\toks0\expandafter{\striprel@x #1}%
\edef\next{\the\toks0}\ifx\next\em@rk\let\next=\endgroup\else\ifx\next\empty%
\else\immediate\write\wfile{\the\toks0}\fi\let\next=\writ@line\fi\next\relax}}
\def\striprel@x#1{} \def\em@rk{\hbox{}}
\def\lref{\begingroup\obeylines\lr@f}
\def\lr@f#1#2{\gdef#1{\ref#1{#2}}\endgroup\unskip}

\def\addref#1{\immediate\write\rfile{\noexpand\item{}#1}} 
\def\startrefs#1{\immediate\openout\rfile=refs.tmp\refno=#1}
\def\xref{\expandafter\xr@f}\def\xr@f#1.{#1}
\def\cite{\expandafter\cxr@f}\def\cxr@f#1.{$^{#1}$}
\def\xcite{\expandafter\xcxr@f}\def\xcxr@f#1.{{#1}}
\def\cites#1{\count255=1$^{\r@fs #1{\hbox{}}}$}
\def\r@fs#1{\ifx\und@fined#1\message{reflabel \string#1 is undefined.}%
\nref#1{need to supply reference \string#1.}\fi%
\vphantom{\hphantom{#1}}\edef\next{#1}\ifx\next\em@rk\def\next{}%
\else\ifx\next#1\ifodd\count255\relax\xref#1\count255=0\fi%
\else#1\count255=1\fi\let\next=\r@fs\fi\next}
\newwrite\lfile
{\escapechar-1\xdef\pctsign{\string\%}\xdef\leftbracket{\string\{}
\xdef\rightbracket{\string\}}\xdef\numbersign{\string\#}}

\def\writestop{\def\writestoppt{\immediate\write\lfile{\string\pageno%
\the\pageno\string\startrefs\leftbracket\the\refno\rightbracket%
\string\def\string\secsym\leftbracket\secsym\rightbracket%
\string\secno\the\secno\string\meqno\the\meqno}\immediate\closeout\lfile}}
\def\writestoppt{}\def\writedef#1{}
\def\seclab#1{\xdef #1{\the\secno}\writedef{#1\leftbracket#1}\wrlabeL{#1=#1}}
\def\subseclab#1{\xdef #1{\secsym\the\subsecno}%
\writedef{#1\leftbracket#1}\wrlabeL{#1=#1}}
\newwrite\tfile \def\writetoca#1{}
\def\leaderfill{\leaders\hbox to 1em{\hss.\hss}\hfill}
\def\writetoc{\immediate\openout\tfile=toc.tmp
   \def\writetoca##1{{\edef\next{\write\tfile{\noindent ##1
   \string\leaderfill {\noexpand\number\pageno} \par}}\next}}}

%
\catcode`\@=12 
%
%
\font\abssl=cmsl10 scaled 833
\font\absrm=cmr10 scaled 833 \font\absrms=cmr7 scaled  833
\font\absrmss=cmr5 scaled  833 \font\absi=cmmi10 scaled  833
\font\absis=cmmi7 scaled  833 \font\absiss=cmmi5 scaled  833
\font\abssy=cmsy10 scaled  833 \font\abssys=cmsy7 scaled  833
\font\abssyss=cmsy5 scaled  833 \font\absbf=cmbx10 scaled 833
\skewchar\absi='177 \skewchar\absis='177 \skewchar\absiss='177
\skewchar\abssy='60 \skewchar\abssys='60 \skewchar\abssyss='60
\def\abstractfont{\def\rm{\fam0\absrm}
\textfont0=\absrm \scriptfont0=\absrms \scriptscriptfont0=\absrmss
\textfont1=\absi \scriptfont1=\absis \scriptscriptfont1=\absiss
\textfont2=\abssy \scriptfont2=\abssys \scriptscriptfont2=\abssyss
\textfont\itfam=\absi \def\it{\fam\itfam\absi}
\textfont\slfam=\abssl \def\sl{\fam\slfam\abssl}
\textfont\bffam=\absbf \def\bf{\fam\bffam\absbf}\rm}
\font\ftsl=cmsl10 scaled 833
\font\ftrm=cmr10 scaled 833 \font\ftrms=cmr7 scaled  833
\font\ftrmss=cmr5 scaled  833 \font\fti=cmmi10 scaled  833
\font\ftis=cmmi7 scaled  833 \font\ftiss=cmmi5 scaled  833
\font\ftsy=cmsy10 scaled  833 \font\ftsys=cmsy7 scaled  833
\font\ftsyss=cmsy5 scaled  833 \font\ftbf=cmbx10 scaled 833
\skewchar\fti='177 \skewchar\ftis='177 \skewchar\ftiss='177
\skewchar\ftsy='60 \skewchar\ftsys='60 \skewchar\ftsyss='60
\def\footnotefont{\def\rm{\fam0\ftrm}
\textfont0=\ftrm \scriptfont0=\ftrms \scriptscriptfont0=\ftrmss
\textfont1=\fti \scriptfont1=\ftis \scriptscriptfont1=\ftiss
\textfont2=\ftsy \scriptfont2=\ftsys \scriptscriptfont2=\ftsyss
\textfont\itfam=\fti \def\it{\fam\itfam\fti}%
\textfont\slfam=\ftsl \def\sl{\fam\slfam\ftsl}%
\textfont\bffam=\ftbf \def\bf{\fam\bffam\ftbf}\rm}
\font\ninerm=cmr9 \font\sixrm=cmr6 \font\ninei=cmmi9 \font\sixi=cmmi6
\font\ninesy=cmsy9 \font\sixsy=cmsy6 \font\ninebf=cmbx9
\font\nineit=cmti9 \font\ninesl=cmsl9 \skewchar\ninei='177
\skewchar\sixi='177 \skewchar\ninesy='60 \skewchar\sixsy='60
\def\ninepoint{\def\rm{\fam0\ninerm}
\textfont0=\ninerm \scriptfont0=\sixrm \scriptscriptfont0=\fiverm
\textfont1=\ninei \scriptfont1=\sixi \scriptscriptfont1=\fivei
\textfont2=\ninesy \scriptfont2=\sixsy \scriptscriptfont2=\fivesy
\textfont\itfam=\ninei \def\it{\fam\itfam\nineit}\def\sl{\fam\slfam\ninesl}%
\textfont\bffam=\ninebf \def\bf{\fam\bffam\ninebf}\rm}
%
%

\vsize=7.0truein
\hsize=4.7truein
\baselineskip 12truept plus 0.5truept minus 0.5truept
\hoffset=0.5truein
\voffset=0.5truein

\def\gsim{\mathrel{\rlap{\lower4pt\hbox{\hskip1pt$\sim$}}
    \raise1pt\hbox{$>$}}}         

\def\frac#1#2{{{#1}\over {#2}}}

\def\GeV{{\rm GeV}}
\def\MS{\hbox{$\overline{\rm MS}$}}

\catcode`@=11 
\def\slash#1{\mathord{\mathpalette\c@ncel#1}}
 \def\c@ncel#1#2{\ooalign{$\hfil#1\mkern1mu/\hfil$\crcr$#1#2$}}
\def\lsim{\mathrel{\mathpalette\@versim<}}
\def\gsim{\mathrel{\mathpalette\@versim>}}
 \def\@versim#1#2{\lower0.2ex\vbox{\baselineskip\z@skip\lineskip\z@skip
       \lineskiplimit\z@\ialign{$\m@th#1\hfil##$\crcr#2\crcr\sim\crcr}}}
\catcode`@=12 

\def\PR{{\it Phys.~Rev.~}}

\def\NP{{\it Nucl.~Phys.~}}
\def\NPBPS{{\it Nucl.~Phys.~B (Proc.~Suppl.)~}}
\def\PL{{\it Phys.~Lett.~}}

\def\APP{{\it Acta.~Phys.~Pol.~}}
\def\vol#1{{\bf #1}}\def\vyp#1#2#3{\vol{#1}, #3 (#2)}


\tolerance=10000
\hfuzz=5pt
\pageno=0\nopagenumbers\tolerance=10000\hfuzz=5pt
\line{\hfill {\tt hep-ph/9607289}}
\line{\hfill Edinburgh 96/10}
\line{\hfill DFTT 36/96}
\vskip 24pt
\centerline{\bf INCLUSIVE MEASUREMENT OF THE}
\centerline{\bf STRONG COUPLING AT HERA}
\vskip 36pt\centerline{Richard D. Ball\footnote*{\footnotefont
Royal Society University Research Fellow}}
\vskip 12pt
\centerline{\it Department of Physics and Astronomy}
\centerline{\it University of Edinburgh, EH9 3JZ, Scotland}
\vskip 18pt
\centerline{and}
\vskip 18pt
\centerline{Stefano Forte}
\vskip 12pt
\centerline{\it INFN, Sezione di Torino}
\centerline{\it via P. Giuria 1, I-10125 Torino, Italia}
\vskip 48pt
{\narrower\baselineskip 10pt
\centerline{\bf Abstract}
\medskip\noindent
We describe the measurement of the strong coupling $\alpha_s$
from data on inclusive DIS at high energies. We present new results
using the 1994 data from H1, and confirm directly the expected
running of $\alpha_s$.
\smallskip}
\bigskip
\centerline{Talk given at {\it DIS96}, Rome, April 1996}
\medskip
\centerline{\it to be published in the proceedings}
\vskip 55pt
\line{June 1996\hfill}
\eject
\footline={\hss\tenrm\folio\hss}
\centerline{\bf INCLUSIVE MEASUREMENT OF THE}
\centerline{\bf STRONG COUPLING AT HERA}
\bigskip\bigskip
{\ninepoint
\centerline{RICHARD D.~BALL}
\smallskip
\centerline{\it Department of Physics and Astronomy,}
\centerline{\it University of Edinburgh, EH9 3JZ, Scotland}
\smallskip
\centerline{and}
\smallskip
\centerline{STEFANO~FORTE}
\smallskip
\centerline{\it INFN, Sezione di Torino}
\centerline{\it via P. Giuria 1, I-10125 Torino, Italia}
}
\bigskip
{\abstractfont\baselineskip 9 pt
\advance\leftskip by 36truept\advance\rightskip by 36truept\noindent
We describe the measurement of the strong coupling $\alpha_s$
from data on inclusive DIS at high energies. We present new results
using the 1994 data from H1, and confirm directly the expected
running of $\alpha_s$.
\smallskip}

\baselineskip 12pt plus 0.5pt minus 0.5pt
\bigskip\bigskip
\goodbreak


\nref\VM{M.~Virchaux and A.~Milsztajn, \PL\vyp{B274}{1992}{221}.}
\nref\DGPTWZ{A.~De~R\'ujula et al., \PR\vyp{D10}{1974}{1649}}
\nref\DAS{R.D.~Ball and S.~Forte, \PL\vyp{B335}{1994}{77}.}
\nref\test{R.D.~Ball and S.~Forte, \PL\vyp{B336}{1994}{77}.}
\nref\paris{G.~Altarelli, summary talk at {\it DIS 95}, Paris, April 1995.}
\nref\Eisele{F.~Eisele, summary talk at the
Europhysics Conference, Brussels, August 1995.}
\nref\zakopane{S.~Forte and R.D.~Ball, \APP\vyp{B26}{1995}{2097}.}
\nref\Hone{H1 Collaboration, \NP\vyp{B439}{1995}{471}.}
\nref\mk{M.~Klein, these proceedings.}
\nref\sf{S.~Forte and R.D.~Ball, {\tt hep-ph/9607291}.}
\nref\Mont{R.D.~Ball and S.~Forte, \NPBPS\vyp{39B,C}{1995}{25}.}
\nref\summ{R.D.~Ball and S.~Forte, \PL\vyp{B351}{1995}{313}.}
\nref\alphas{R.D.~Ball and S.~Forte, \PL\vyp{B358}{1995}{365}.}
\nref\mrs{A.D.~Martin et al, {\tt hep-ph/9506423}.}
\nref\gmrs{E.W.N.~Glover et al, {\tt hep-ph/9603327}.}
\nref\wkt{W-K.~Tung, these proceedings.}
\nref\rgr{R.G.~Roberts, these proceedings.}

\noindent

Determining the strong coupling $\alpha_s$ from fixed target inclusive deep
inelastic scattering\cite\VM\ requires very precise
measurements of structure functions at large $x$ over a wide range
of $Q^2$ in order to extract the relatively small violations
of Bjorken scaling. In colliding beam deep inelastic scattering
experiments, such as are now being performed at HERA, structure
functions may be measured at small $x$, where Bjorken scaling
violations are much stronger. Indeed the steep rise in $F_2(x,Q^2)$
as $1/x$ and $Q^2$ both increase is driven essentially by the
three-gluon vertex\cite\DGPTWZ, resulting asymptotically in a new double
scaling behaviour\cite\DAS. The predicted slope of the rise
in $F_2$ has been confirmed very precisely using the recent HERA
data\cites{\test-\Eisele}, and two loop corrections are now
discernible\cites{\zakopane-\sf}. Since the three gluon vertex is
itself directly proportional to $\alpha_s$, the success of double
scaling immediately suggests\cite\DAS\ that $F_2(x,Q^2)$ must be
rather sensitive to the value of $\alpha_s$ in the double scaling
region, and indeed this expectation is confirmed by simple double
scaling fits to the data which include $\alpha_s$ as a free
parameter\cites{\alphas,\mk}.

Of course a proper determination\cite\alphas\ of $\alpha_s$ from
$F_2$ at small $x$ requires a full two loop calculation which
matches on to all the other data (and in particular structure
function and prompt photon data) at large $x$. Care must also be taken to
disentangle the effects of higher twists and higher logarithms
(of $1/x$): fortunately this turns out to be rather easier than
at large $x$, where the effects of both higher twists and
infrared logarithms (of $1-x$) may become very significant.
Here we will briefly review how such a small $x$ determination
may be performed, and then present some preliminary results
obtained using the most recent data\cite\Hone\ from the H1 collaboration.

\topinsert
\vskip-2.5truecm
\vbox{\hbox{\hskip-0.5truecm
\hfil\epsfxsize=6.5truecm\epsfbox{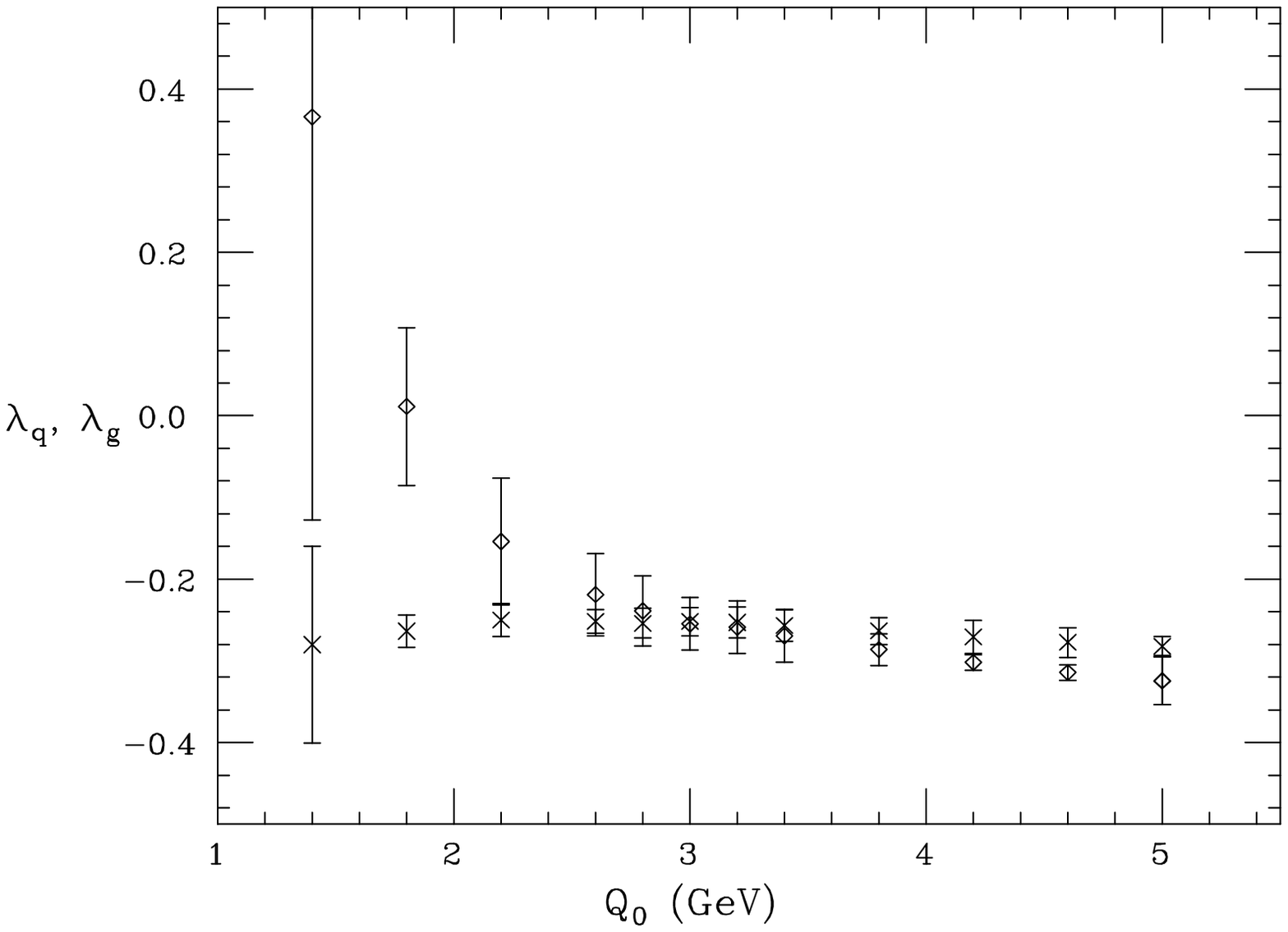}
\hskip -0.5truecm
\epsfxsize=6.5truecm\epsfbox{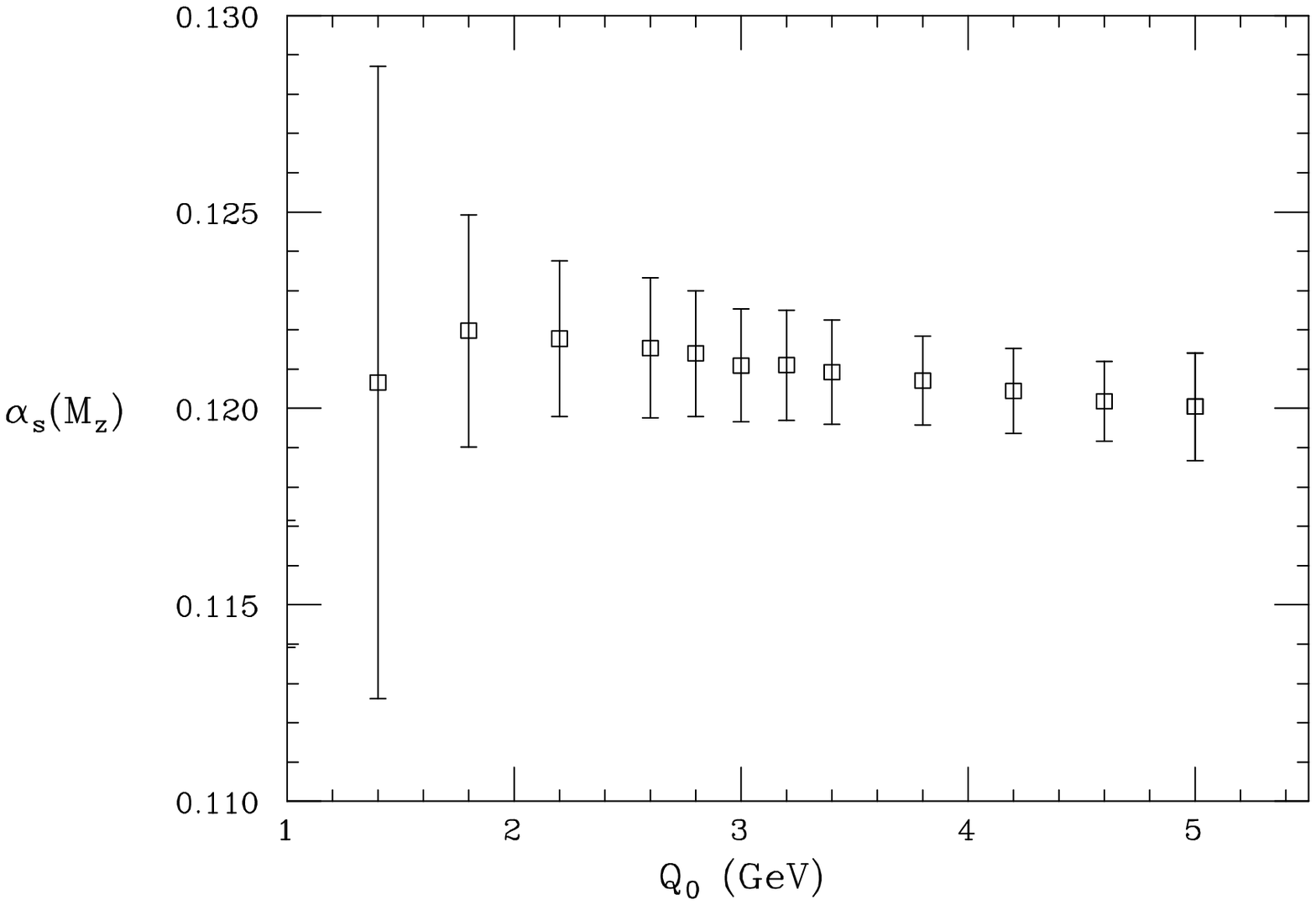}\hfil}
\vskip -2.5truecm
\bigskip\noindent{\footnotefont\baselineskip6pt\narrower
Figure 1: The three fitted parameters
$(\lambda_q,\lambda_g,\alpha_s(m_Z))$ for a range of values of $Q_0$.
The distribution is fitted and evolved in \MS.\vskip0.1truecm
}}
\vskip -0.2truecm
\vbox{\hbox{
\hfil\epsfxsize=3.6truecm\epsfbox{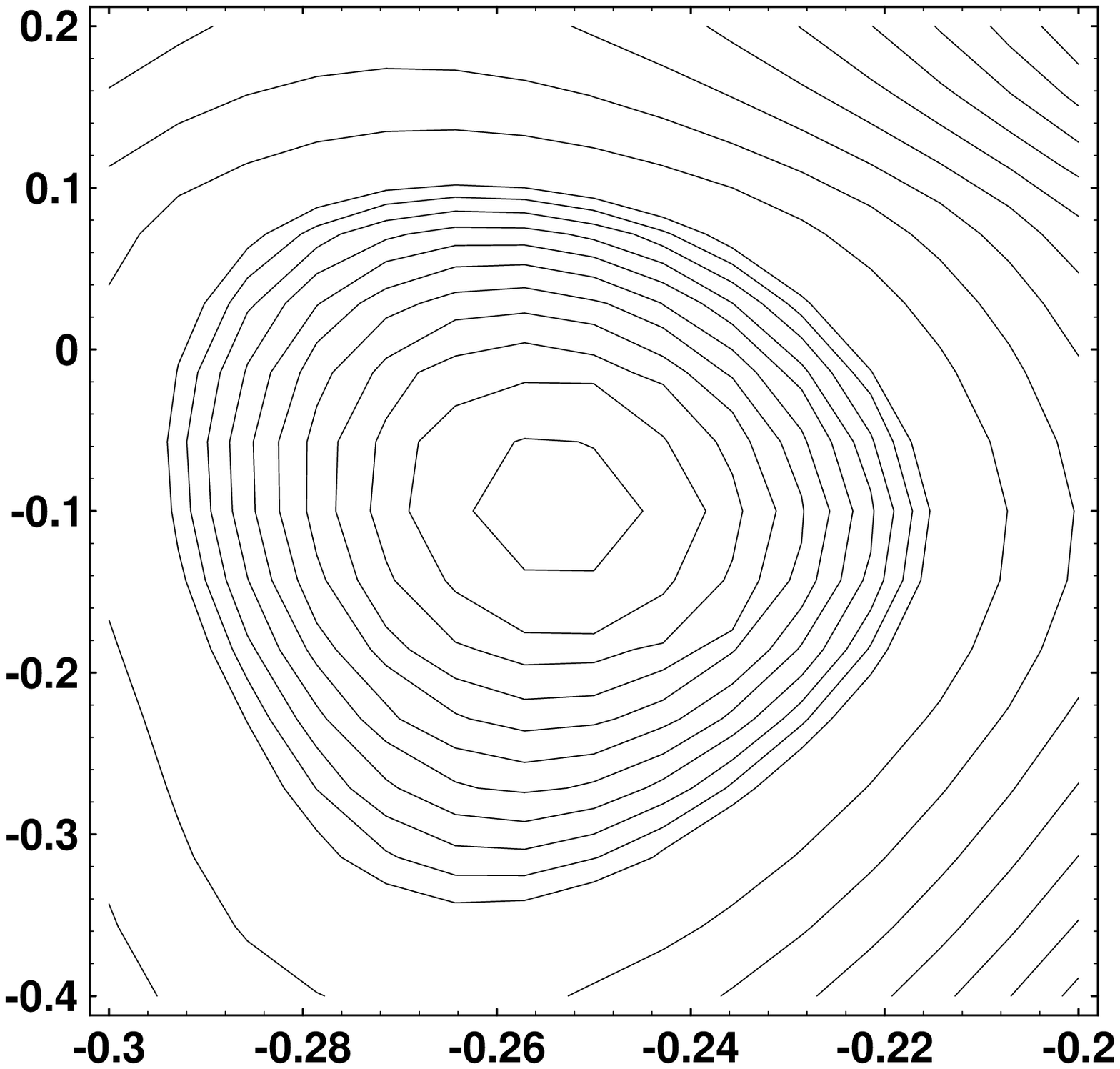}
\hskip 0.3truecm
\epsfxsize=3.6truecm\epsfbox{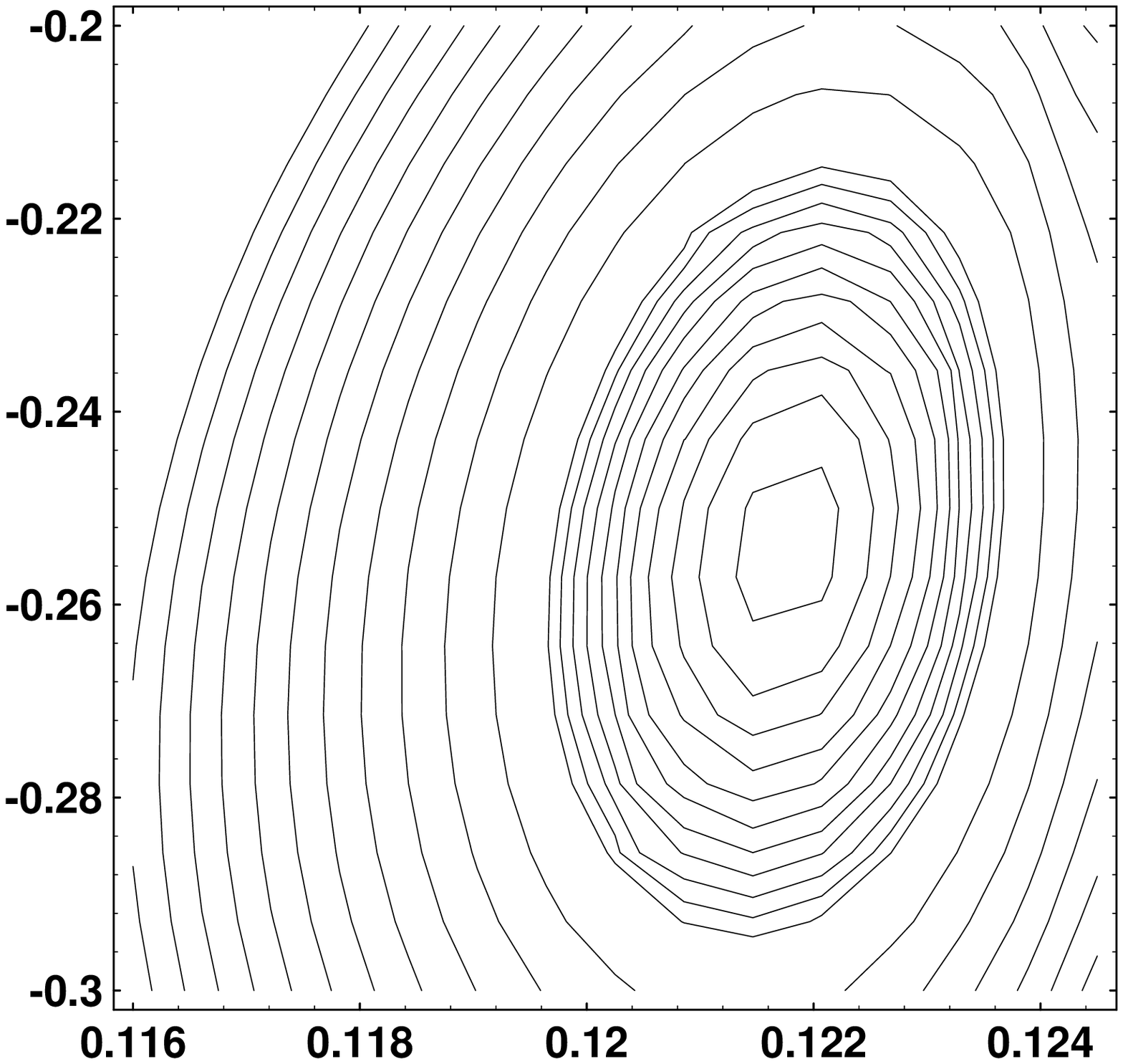}
\hskip 0.3truecm
\epsfxsize=3.6truecm\epsfbox{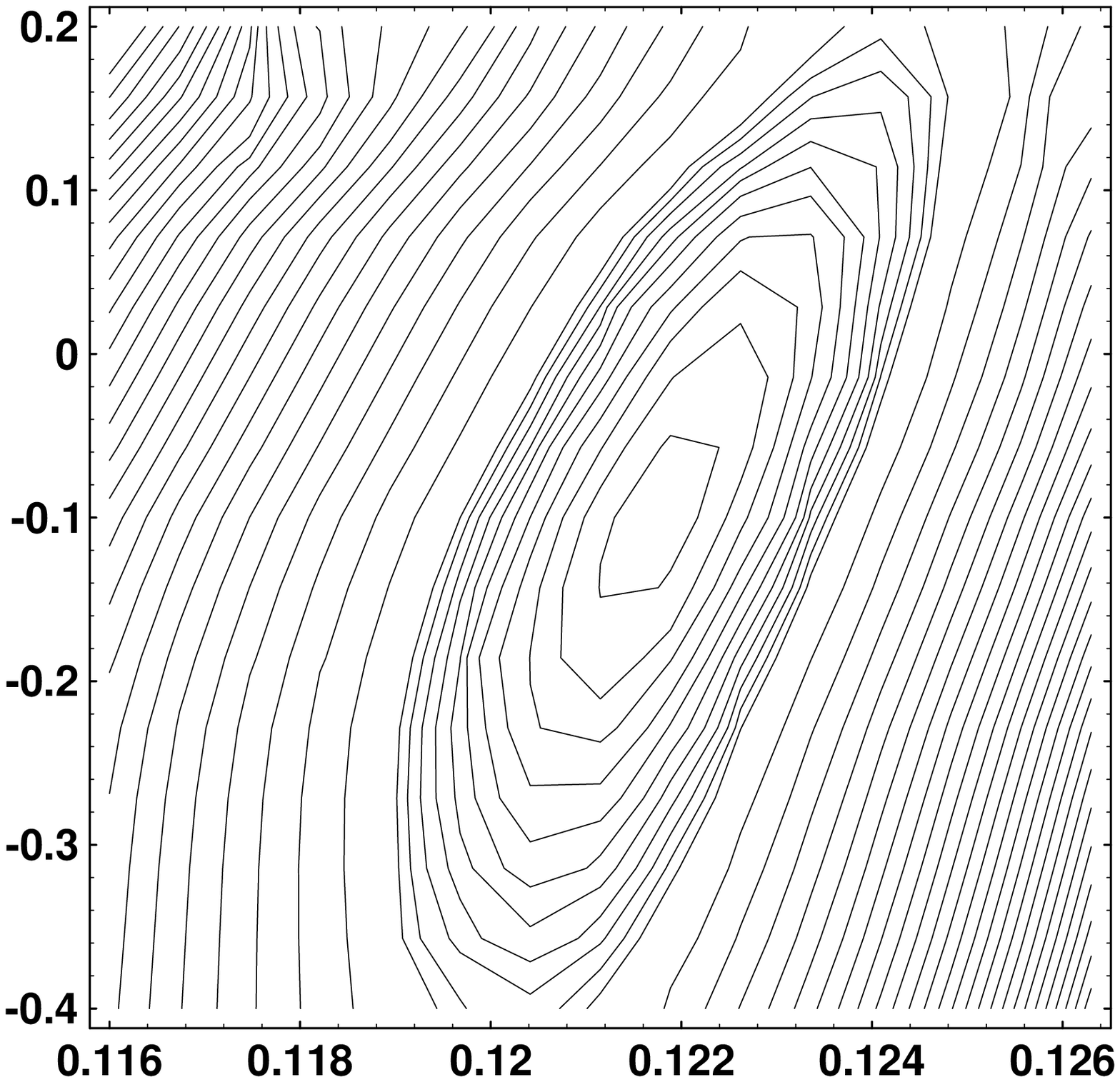}
\hfil}
\vskip-1.0truecm
\bigskip\noindent{\footnotefont\baselineskip6pt\narrower
Figure 2: Contour plots of $\chi^2$ in the three orthogonal planes
$(\lambda_q,\lambda_g)$, $(\alpha_s(m_Z),\lambda_q)$ and
$(\alpha_s(m_Z),\lambda_g)$ through the global minimum. The first eleven
contours are at intervals of one unit, while those thereafter are at
intervals of five units. The starting distribution is in \MS\ at
$Q_0=2\GeV$. To compute the $\chi^2$ statistical and systematic errors
are combined in quadrature, and the normalization uncertainties
included: at the minimum $\chi^2=80$ for $166$ degrees of freedom,
the normalizations are within $1\%$ of the experimental values.
\vskip0.1truecm}}
\medskip
\endinsert

Our basic procedure\cites{\Mont,\alphas} is to take a set of globally fitted
parton distributions, for example those of CTEQ\cite\wkt\ or
MRS\cite\rgr, evolve them (using a two loop evolution code) to some
new starting scale $Q_0$, and there cut off the tails of the sea and
gluon distributions, replacing them with new tails
$xq\sim x^{\lambda_q}$, $xg\sim x^{\lambda_g}$ for $x\lsim
0.01$.\footnote{*}
{\footnotefont\baselineskip6pt
In practice this is achieved by refitting
the large $x$ distributions
at $Q_0$ with $\lambda_q$ and $\lambda_g$ kept fixed.\vskip0.1truecm}
These new distributions, together with the valence distributions,
are evolved up (or
down) to the HERA data, again at two loops, using a particular
value of $\alpha_s(m_Z)$. The three parameters
$(\lambda_q,\lambda_g,\alpha_s(m_Z))$ are then adjusted to
minimise the $\chi^2$ to the data. The final distributions
are by construction a reasonable fit to all the data with $x\gsim 0.01$
used to determine the original input
distributions, since QCD evolution is causal in $x$. However the
shape of the distributions in the HERA region is largely independent
of the details of the input, since asymptotically it takes a universal
(double scaling) form which depends rather sensitively on the
value of $\alpha_s$.\cite\DAS\  Higher
order effects due to choosing large $x$ distributions which have been
fitted using different values of $\alpha_s$ should in principle
also be included. In this sense our procedure is similar to
that used to determine $\alpha_s$ from, for example, the $2+1$ jet rate at
HERA, or the inclusive jet rate at the Tevatron.

\topinsert
\vskip-2.5truecm
\vbox{\hbox{\hskip2truecm
\hfil\epsfxsize=8truecm\epsfbox{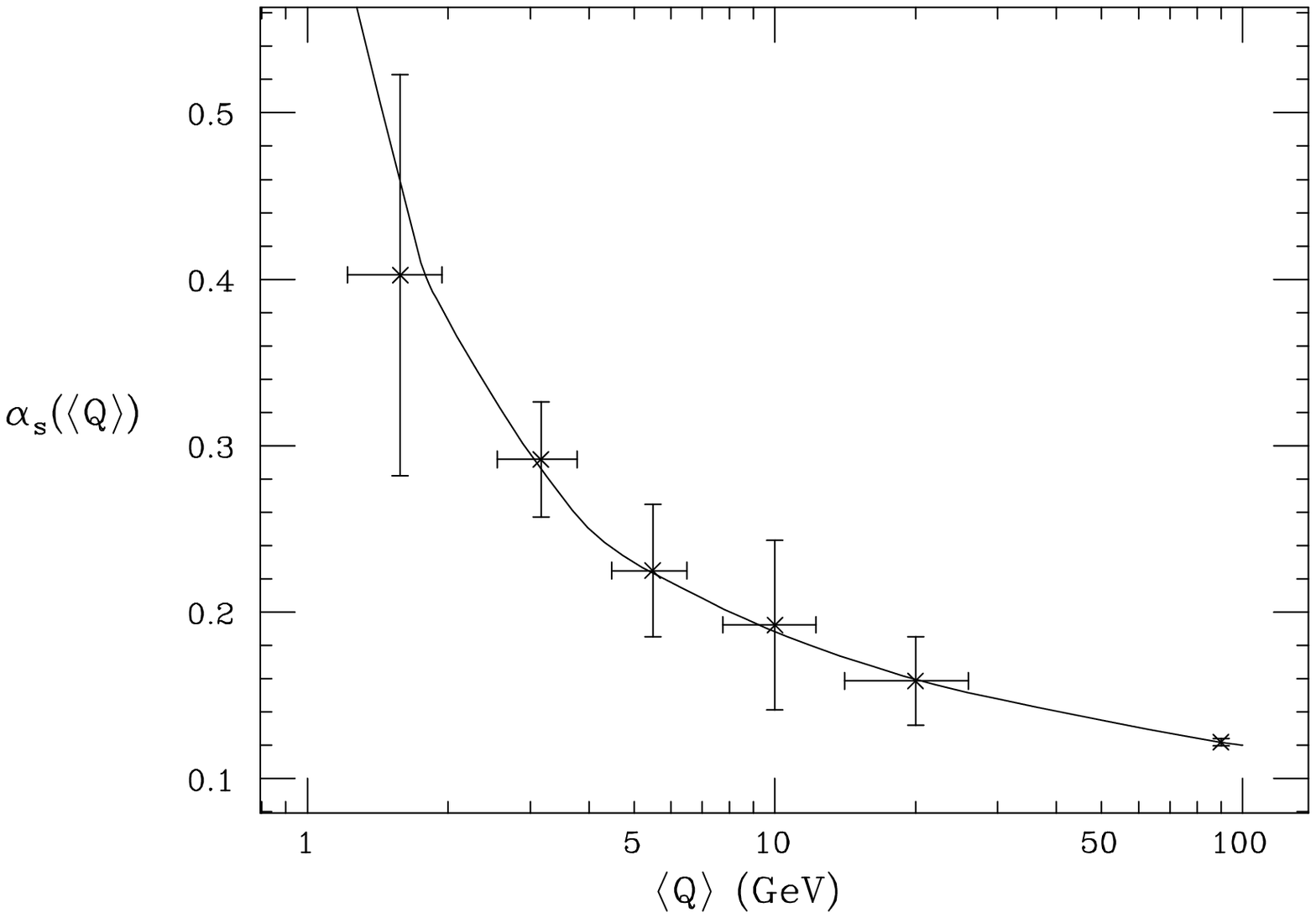}\hfil}
\vskip-3.0truecm
\bigskip\noindent{\footnotefont\baselineskip6pt\narrower
Figure 3: The strong coupling measured independently at five different
scales. The horizontal bars indicate the width of each bin.
The curve shows the theoretical two loop running when
$\alpha_s(m_Z)=0.122$ (the sixth point). The points are
over-correlated due to correlated systematics.\vskip0.1truecm}}
\medskip
\endinsert

The results of such fits to the H1 1994 data\cite\Hone\ for a range
of values of $Q_0$ are presented
in fig.~1. The stability of the fitted value of $\alpha_s$ in the
region where the $\chi^2$ is lowest (2~$\GeV\lsim Q_0\lsim4$~GeV) is a
useful test of the effectiveness of the parametrization employed at
$Q_0$. More information about the shape of the minimum in the $\chi^2$ at
a particular value of $Q_0$ is displayed in the contour plots fig.~2.
The third plot shows that although the shape of the inferred
gluon distribution and $\alpha_s$ are indeed correlated, this
does not lead to a large uncertainty in $\alpha_s$ because of the
close relation between the quark and the gluon shown in the
first plot: very steeply rising
or falling gluon distributions at 2~$\GeV$ are incompatible with the
double scaling seen in the H1 data. This explains why it is possible to
determine $\alpha_s$ at small $x$ even though the gluon distribution
is large there.

Our preliminary result using the H1 1994 data is
$\alpha_s(m_Z)=0.122\pm 0.004 {\rm (exp)}$.
The theoretical error has not yet been determined, but should be a
little less than that found\cite\alphas\ in
our analysis of the 1993 data, partly because the data themselves now
severely limit the allowed size of corrections due to higher order
logarithms of $1/x$.\cite\sf\ The error due to higher twist corrections
seems also to be very small, since their size is again limited by the
data.\cite\sf\ Furthermore the allowed range of
variation of renormalization and factorization scales can now be
limited by requiring that the fit to the data is not unduly worsened.
A more complete analysis will be made when all the ZEUS 1994 data have
been published.

\topinsert
\vskip-3truecm
\vbox{\hskip 5.5truecm\hbox{\hskip 1.5 truecm
\epsfxsize=9truecm\epsfbox{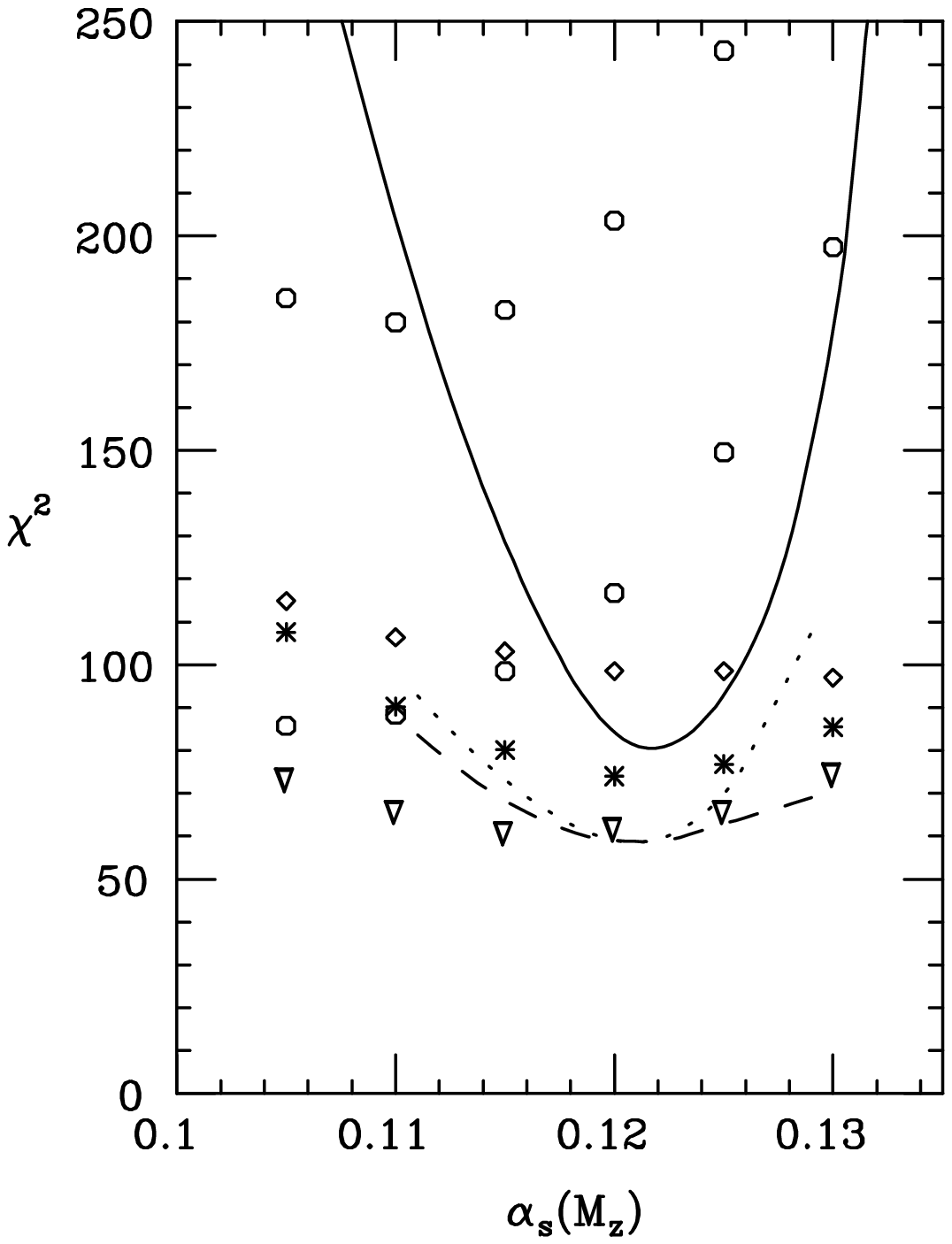}}
\vskip-2.5truecm
\bigskip\noindent{\footnotefont\baselineskip6pt\narrower
Figure 4: Contributions to the total $\chi^2$ from various structure
function data sets included
in global fits at six different values of $\alpha_s(M_Z)$, adapted from
ref.\xcite\mrs. The data are from BCDMS $F_2$ (open squares, 142
points), NMC $F_2$ (open diamonds, 74 points),
CCFR $F_2$ (open circles, 80 points), CCFR $F_3$
(open triangles, 80 points), and HERA $F_2$ (asterisks, 93 points from
H1 $+$ 56 from ZEUS, both from the 1993 run). The curves are also from
fits to HERA $F_2$ data: the dotted ($\lambda_q=\lambda_g$) and dashed
($\lambda_q\neq\lambda_g$) curves using 1993 data (122 points) are
from the analysis in ref.\xcite\alphas, while the solid curve is from
this analysis (H1 1994 data, 169 points). In all these curves
$\alpha_s(M_Z)$ in the large $x$ input distribution was kept fixed.
\vskip0.1truecm
}}
\medskip
\endinsert

By only including some of the data in the fit, over a limited range of
$Q^2$, we can actually perform independent measurements of $\alpha_s$
at different scales. The results of such a procedure are displayed in
fig.~3. It is gratifying to observe directly the running of $\alpha_s$
in a single experiment. Of course we already knew that this had to
work: the slope of the rise of $F_2$ is in itself a direct measurement
of the first coefficient of the $\beta$-function.\cite\test

It is important to emphasise that our determination uses only HERA
data directly: it is not a global fit. Some of the
advantages and disadvantages of the two techniques may be seen by
comparing our results to those of the subsequent MRS global fits described in
ref.\xcite\mrs\ (see fig.~4), or indeed to very similar results from
CTEQ.\cite\wkt\ In a global fit the $\chi^2$ for each data set
are in general simply added together, irrespective of the fact that different
experimental collaborations may treat their errors rather differently.
The data from BCDMS then dominate the determination of $\alpha_s$
not because there are many more points, but because the $\chi^2$ per
point turns out to be relatively large (a better fit to these
data, with more parameters, was obtained in ref.\xcite\VM). Indeed
the minima in the
$\chi^2$ of the NMC and CCFR $F_2$ data actually disappear in the MRS
global fit. The minimum in the 1993 HERA data is uncorrupted, however, and
is indeed consistent both in position and width with the direct
analysis of ref.\xcite\alphas: this is to be expected as the shape of
$F_2$ at small $x$ is largely independent of the detailed structure at
large $x$.\cite\DAS\ Note that the global analysis\cite\mrs\ has
$\lambda_q=\lambda_g$: relaxing this constraint softens the minimum
but does not eliminate it. The present analysis of the 1994 data
shows that in the future HERA structure function data will begin to play an
increasingly significant role in the global fitting of $\alpha_s$ (although our
solid curve in fig.~4 will presumably not fall quite so steeply
once $\alpha_s$ is varied in the input distribution). It is interesting
that the relatively high value of $\alpha_s$ favoured by HERA is also
found in analyses of the CDF and D0 inclusive jet data.\cites{\gmrs,\wkt,\rgr}

In conclusion, $F_2^p$ at small $x$ and large $Q^2$ is an excellent
place to measure $\alpha_s$: the data is now very precise, the dependence
on $\alpha_s$ is strong, one need only do simple fits with a small number of
parameters, while on the theoretical side uncertainties due to higher
logarithms seem to be unexpectedly small, while higher
twists are truly negligible. Furthermore one can perform a direct test
of the running of $\alpha_s$ at spacelike $Q^2$ in one experiment. We
hope that soon experimentalists will perform their own
determinations, taking properly into account the correlations of their
systematic errors.

\immediate\closeout\rfile\writestoppt
\bigskip
\noindent{{\bf References}}\smallskip{\frenchspacing%
\parindent=20pt
\ninepoint\baselineskip=11pt
\escapechar=` \input refs.tmp\vfill\eject}\nonfrenchspacing

\end